\documentstyle[epsf,12pt]{article}
\newcommand{\gsim}{\mbox{\raisebox{-.3em}{$\stackrel{>}{\sim}$}}}
\newcommand{\lsim}{\mbox{\raisebox{-.3em}{$\stackrel{<}{\sim}$}}}
\renewcommand{\thefootnote}{\fnsymbol{footnote}}

\renewcommand{\cite}[1]{\ref{#1}}

\newcommand{\half}{\frac{1}{2}}
\newcommand{\reflef}{(\ref}
\newcommand{\beq}{\begin{equation}}
\newcommand{\eeq}{\end{equation}}
\newcommand{\beqa}{\begin{eqnarray}}
\newcommand{\eeqa}{\end{eqnarray}}
\newcommand{\bcent}{\begin{center}}
\newcommand{\ecent}{\end{center}}

\begin{document}

%\begin{flushright}
%\bcksl scalar\bcksl qtst3a.tex \today
%\end{flushright}

\baselineskip=0.8cm
\bcent
{\Large\bf Quintessence, scalar-tensor theories and non-Newtonian gravity}\\[2.em]
\baselineskip=0.6cm
Yasunori Fujii\footnote{E-mail address: fujii@handy.n-fukushi.ac.jp}\vspace{.6em}\\
Nihon Fukushi University, Handa, 475-0012\ Japan\\
\ecent
%\mbox{}\\[-1.5em]
\bcent
{\large\bf Abstract}\\[1.em]
\baselineskip=0.5cm
\parbox{13cm}{We discuss some of the issues which we encounter when
we try to invoke the scalar-tensor theories of gravitation as a theoretical basis of quintessence.   One of the advantages of appealing to these theories
 is that they allow us to implement  the scenario of a
``decaying cosmological constant,''  which offers a reasonable
understanding of why the observed upper bound of the cosmological 
constant is smaller than the theoretically natural value by as much 
as 120 orders of magnitude.  	In this context, the scalar
field can be a candidate of quintessence in a broader sense. We find,
however, a serious drawback in the prototype Brans-Dicke model with
$\Lambda$ added; a static universe in the physical conformal frame
which is chosen to have constant particle masses.  We propose a remedy
by modifying the matter coupling of the scalar field taking advantage
of  scale invariance and its breakdown through quantum anomaly.  By
combining this with a conjecture on another cosmological constant
problem coming from the vacuum energy of matter fields,  we expect a
possible link between quintessence and non-Newtonian gravity featuring 
violation of Weak Equivalence Principle and intermediate force range, 
likely within the experimental constraints.  A new prediction is also 
offered on the time-variability of the gravitational constant.
}
\ecent

%\newpage

\renewcommand{\thefootnote}{\arabic{footnote}}

\setcounter{footnote}{0}
\baselineskip=0.6cm

\section{Introduction}

The role of the gravitational scalar field, now widely called quintessence, has been a
subject of extensive studies [\cite{qt1},\cite{qt2}] to understand a small cosmological
``constant," as suggested by recent observations indicating an accelerating universe [\cite{sn}].  Much of the interest
seems to be directed to the question how the dynamics of the scalar
field can simulate observed behaviors.  Some of the
theoretical models, notably the scalar-tensor theories, have been
discussed, on the other hand, as a possible solution of the
``cosmological constant problem [\cite{ccp}]."  In this paper we discuss some
 of the  crucial aspects of the arguments which we may encounter in
this approach.

It seems useful to recognize that a cosmological constant may have two
different origins.  First, in almost any of the theoretical models of
unification, we have  no way to avoid to have $\Lambda$ in the
Lagrangian, with the magnitude typically of the order unity in the
reduced Planckian unit system, with $c=\hbar =M_{\rm p}^{-2}=8\pi G
=1$.  This constant is too large compared with the observed upper
bound, given basically by the critical density, by as much as 120 orders of magnitude.  Secondly, the vacuum energy of matter fields, in the sense of relativistic quantum field theory, plays the same role as the cosmological constant.  This contribution is also too large by somewhere around 60 orders of magnitude.  We call them conveniently the primordial cosmological constant and the vacuum energy, respectively.

We focus upon the scalar-tensor theories of gravitation which have been
discussed from many points of view, sometimes quite different from the
original motivation.  One of the reasons of our interest comes from
the fact that it provides us with an exponential potential of the
scalar field rather naturally as a direct consequence of introducing a
primordial cosmological constant [\cite{ccp}].  It offers a way to put
the discrepancy of 120 orders of magnitude under control without
appealing to  an extreme fine-tuning of parameters.   According to the ``scenario of a decaying cosmological constant," today's $\Lambda$ is small only because our universe is old.  The number comes simply from $t^{-2}\sim 10^{-120}$ for the present age of the universe $t\sim 10{\rm Gy}\sim 10^{60}$ in the reduced Planckian unit system.

The effective cosmological constant, the energy of the scalar field,
falls off, however, in the same way as the ordinary matter density.
For this reason this model does not result in the extra acceleration
of the universe.   Further development is called for to understand the
lower bound as well [\cite{qt1},\cite{qt2},\cite{yf3}].  The extension 
based on the exponential potential still seems promising because it
will inherit the decaying nature of the cosmological constant.  As
another potential advantage, we expect to shed a new light on the
coincidence problem.

The same ``success" does not seem promising, however, for the vacuum
energy contribution.  Moreover, we find that the prototype Brans-Dicke
(BD) model suffers from a serious drawback once $\Lambda$ is
introduced [\cite{yf4}].  As a remedy we proposed a modification in
the scalar-matter coupling taking a risk of violating  Weak
Equivalence Principle (WEP).  We foresee the scalar field to show up
as non-Newtonian gravity, or the fifth force, once we find a way to
solve the vacuum energy problem by a still unknown mechanism.

In Section 2 we sketch the solution of the prototype BD model with the
primordial cosmological constant added.  After explaining how the
model suffers from the drawback, we outline an alternative, the
dilaton model in Section 3, featuring WEP violation.  Section 4
discusses non-Newtonian gravity, which is expected to arise as a
residual effect.  As one of key arguments, we emphasize that the
force-range of the quantized field that mediates a force between local
matter objects can be different from the counterpart of the classical
field which governs the evolution of the universe.  In the final
Section 5 we discuss a modification by allowing the self-coupling of
the scalar field, as well as the predicted time variability of the gravitational constant.

%%%%%%%%%%%%%%%%%%%%%% sect 2 %%%%%%%%%%%%%%%%%%%%%%%%%%%%%%%%%%%%%%%
\section{Prototype BD model with $\Lambda$ added}

Consider the Lagragian for the prototype BD model with $\Lambda$ added [\cite{yf4}]:
\beq
{\cal L}_{\rm BD}=\sqrt{-g}\left( \half\xi\phi^2 R -\epsilon\half g^{\mu\nu}\partial_{\mu}\phi \partial_{\nu}\phi -\Lambda +L_{\rm matter} \right),
\label{qtst-1}
\eeq
where $\xi>0$ is a dimensionless constant related to the original symbol $\omega$ by $4\xi\omega = \epsilon = {\rm Sign}(\omega)$.  Also our $\phi$ is a canonical field related to BD's original field $\varphi$ by $\varphi =(\xi /2)\phi^2$.

It is useful to apply a conformal transformation
\beq
g_{\mu\nu} \rightarrow g_{*\mu\nu} = \Omega^2(x)g_{\mu\nu},
\quad\mbox{with}\quad
\Omega = \xi^{1/2}\phi.
\label{qtst-2}
\eeq
We have thus moved from the original conformal frame (CF) called J frame after Jordan to a new conformal frame (called E frame after Einstein) in which
\reflef{qtst-1}) is re-expressed as
\beq
{\cal L}_{\rm BD}=\sqrt{-g_*}\left( \half R_* -\half
g_{*}^{\mu\nu}\partial_{\mu}\sigma\partial_{\nu}\sigma - V(\sigma)
+L_{*\rm matter} \right),
\label{qtst-3}
\eeq
where the new canonical scalar field $\sigma$ is defined by
\beq
\phi = \xi^{-1/2}e^{\zeta\sigma},\quad\mbox{with}\quad  \zeta^{-2}=6+\epsilon\xi^{-1},
\label{qtst-4}
\eeq
under the condition
\beq
 \epsilon \xi^{-1} > -6.
\label{qtst-5}
\eeq
Also notice that the cosmological constant in J frame acts now as a
potential of the scalar field;
\beq
V(\sigma) = \Lambda \Omega^{-4}= \Lambda e^{-4\zeta\sigma}.
\label{qtst-6}
\eeq

In the spatially flat Robertson-Walker universe the cosmological
equations in E frame are
\beqa
&& 3H_* =\rho_{\sigma}+\rho_{*m}, \label{qtst-7}\\
&& \ddot{\sigma}+3H_*\dot{\sigma} -4\zeta V = \eta_{d}\zeta\rho_{*m}, \label{qtst-8}\\
&& \dot{\rho}_{*m} + ( 4-\eta_{d})H_*\rho_{*m} =-\eta_{d}\zeta\dot{\sigma}\rho_{*m},\label{qtst-9}
\eeqa
where we have assumed the spatially uniform $\sigma$ with $\rho_{\sigma} = (1/2)\dot{\sigma}^2 +V$, and $\eta_{d} =0,1$
for the  radiation- and dust-dominated universe, respectively.  Hereafter
we often attach the symbol $*$ to signify quantities in E frame.

Notice that, unlike in J frame, we have the non-vanishing right-hand
side of  \reflef{qtst-9}) in the dust-dominated
universe.  This corresponds to the geodesic equation for a point particle acquiring a nonzero right-hand side.  This does not imply WEP violation because the extra force is proportional precisely to the mass.

We find an analytic solution [\cite{yf4}]
\beqa
a_*(t_*)&=&t_*^{1/2}, \label{qtst-10}\\
\sigma (t_*)&=&\bar{\sigma} + \zeta^{-1}\half \ln t_*,
\label{qtst-11} \\
\rho_{*m}(t_*)&=&\left( 1-\frac{1}{4}\zeta^{-2} \right) t_*^{-2}\times \left\{\begin{array}{l} \frac{3}{4}, \\[.5em]
       1,  
\end{array}\right.
\label{qtst-12} \\[.8em]
\rho_{\sigma}(t_*)&=& t_*^{-2}\times \left\{
\begin{array}{l}
\frac{3}{16}\zeta^{-2}, \\[.5em]
\frac{1}{4}\left( -1+\zeta^{-2} \right),
\end{array}
\right.
\label{qtst-13}
\eeqa
where $\bar{\sigma}$ is defined by 
\beq
\Lambda e^{ -4\zeta\bar{\sigma}} =  \left\{
\begin{array}{l}
\frac{1}{16}\zeta^{-2},\\[.6em]
\frac{1}{8}\left( \zeta^{-2}-2 \right).
\end{array}\right.
\label{qtst-14}
\eeq
The upper and lower lines in \reflef{qtst-12})-\reflef{qtst-14}) are for the radiation- and dust-dominated universe, respectively.    We point out that \reflef{qtst-10})-\reflef{qtst-14}) represent an attractor solution realized asymptotically.  According to \reflef{qtst-13}), the effective cosmological constant $\Lambda_{\rm eff}= \rho_{\sigma}$ falls off like $\sim t_*^{-2}$, implementing the scenario of a decaying cosmological constant.

We find many differences from the solution obtained without $\Lambda$ [\cite{bd}].  The scalar field continues to increase because the potential \reflef{qtst-6}) keeps driving $\sigma$ toward infinity, whereas $\sigma$ comes eventually to rest in the radiation-dominated universe if $\Lambda =0$.  Also the presence of $\rho_{\sigma}$ allows $\rho_{*m}$ to be positive only for $\zeta^2 > 1/4$, from which follows $\epsilon = -1$, but still with a positive energy for the ``diagonalized" $\sigma$.  Rather unexpectedly, the scale factor grows in the same way both in the radiation- and dust-dominated universe.

The above condition $\zeta^2 > 1/4$ from \reflef{qtst-12}) is in
contradiction with the widely accepted constraint $4\zeta^2 \lsim
10^{-3}$, or often expressed as $\omega \gsim 10^3$, obtained from the
solar-system experiments.  The constraint might be avoided if the
scalar field acquires a nonzero mass giving a force-range shorter than
 the size of the solar system.

BD chose $\phi$ to be absent in $L_m$ in J frame to ensure WEP to hold [\cite{bd}].  In E frame, however, mass of any particle depends on $\sigma$;
\beq
m_*(\sigma) =m_0 \Omega^{-1} =m_0 e^{-\zeta\sigma},
\label{qtst-15}
\eeq
where $m_0$ is a constant mass in J frame.  By substituting from
\reflef{qtst-11}), we find $m_*$ to vary like $\sim t_*^{-1/2}$.
Corresponding to the situation in which we analyze primordial
nucleosynthesis, for example, based on quantum mechanics with particle
masses taken as constant, we should select J frame as a physical CF
instead of E frame.  The exessive time dependence may be found also in 
  a longer time span; today's quark mass $m_{q*{\rm p}}\approx 5 {\rm
MeV}\sim 2\times 10^{-21}$, for example, is extrapolated back to $t_*
=1$ giving as large a value as $10^9$.

The behavior of the scale factor $a(t)$ in J frame can be obtained most easily 
 by using the relations 
\beq
dt_* = \Omega dt, \quad\mbox{and}\quad a_* =\Omega a.
\label{qtst-15a}
\eeq
We find $a(t)={\rm const}$.  The universe looks {\em static} in
{\em both} of the radiation- and dust-dominated universe.  The same
result can also be obtained directly in J frame.

It might be useful to discuss what the underlying reason of this
result is at least for the radiation-dominated universe.  First the result \reflef{qtst-10}) is a direct consequence of \reflef{qtst-9}) in which the right-hand side vanishes because $\sigma$ has no source in the radiation-dominated universe.  Secondly, \reflef{qtst-7}) implies that the potential $V(\sigma)\sim \Omega^{-4}$ must behave like $t_*^{-2}$, from which follows $m_* \sim \Omega^{-1}\sim t*^{-1/2}$.  We now find that $a_*(t_*)$ and $m_*^{-1}$ grows in the same way.  The meter stick provided by $m_*^{-1}$ expands in the same way as the universe.  Using this kind of meter stick corresponds exactly to living in J frame.

This conclusion may not be final, because it depends on the
simplest choice of the nonminimal coupling, as well as the assumption
of no self-coupling of $\phi$.  It seems still devastating because the
simplest choice is so remote from what we expect from the standard
cosmology.  We find it far from easy to understand why the universe expands 
in the manner of the standard model.  It might be worth looking for a
remedy from quite a different point of view, still on a simple
theoretical basis.

%%%%%%%%%%%%%%%%%%%%%%%%%%%%%%%%%% sect 3 %%%%%%%%%%%%%%%%%%%%%%%%%%%%%%%%%%
\section{Dilaton model}

A possible alternative might be found by favoring E frame in which we have the standard result $a_*(t_*)=t_*^{1/2}$ for the radiation-dominated universe.  Let us expect that the E frame mass term is given by
\beq
{\cal L}_{mq} = -m_{q\dagger}\sqrt{-g_*}\bar{q}_*q_*,
\label{qtst-16}
\eeq
where $m_{q\dagger}$ is a {\em constant} mass of the quark, for example.  Obviously, this can be conformally transformed back to the Yukawa interaction in J frame;
\beq
{\cal L}_{mq} = -\xi^{1/2}m_{q\dagger}\sqrt{-g}\bar{q}q\phi.
\label{qtst-17}
\eeq
Notice that the coupling constant is dimensionless, as verified by
re-installing $M_{\rm P}^{-1}$.  Allowing the scalar field to enter the
matter Lagrangian should, however,  endanger WEP, but without spoiling Equivalence Principle at the more fundamental level stating that tangential space to curved spacetime should be Minkowskian.  Also violation of WEP as a phenomenological law can be tolerated within the constraint obtained from the fifth-force-type experiments [\cite{fsb}].

This favorable result is lost, however, once the effect of
interactions among matter fields is taken into account.  The QCD
calculation, with the help of dimensional regularization, corresponding to the 1-loop diagram in Fig. 1 yields the linear term [\cite{yf4}]
\beq
{\cal L}_{mq1} = \sqrt{-g_*}\zeta_q m_{q\dagger}\bar{q}_*q_*\sigma,
\label{qtst-18}
\eeq
where
\beq
\zeta_q =\zeta \frac{5\alpha_s}{\pi}\approx 0.3 \zeta,
\label{qtst-19}
\eeq
with $\alpha_s\approx 0.2$, the QCD counterpart of the fine-structure
constant.  WEP violation is explicit because $\zeta_q$ depends on $\alpha_s$ which is specific to the quark.

The coupling strength indicated by \reflef{qtst-19}) is not
considerably weaker than $\zeta$ in the linear term in
\reflef{qtst-15}).  It  still seems sufficient to suppress  particle
masses at the earliest universe to a ``reasonable'' size.  Even
$M_{*sb}\sim 1 {\rm TeV}\sim 4\times 10^{-16}$ for the mass scale of
supersymmetry breaking remains $\sim 10^{-6}$ at $t_* =1$, if
\reflef{qtst-18}) is justified to be exponentiated for large $\sigma$.
We need, however, a detailed analysis on whether the many-loop
calculation for terms higher order in $\sigma$ would result in the
exponential function.  A different asymptotic behavior might emerge.

From a more realistic point of view, however, we may focus on the nucleon mass.  By considering that the quark mass content of a nucleon is relatively small, $\sim 60{\rm MeV}$, we estimate
\beq
\zeta_N \approx 0.02\zeta,
\label{qtst-20}
\eeq
suggesting that E frame can be a physical CF to a good approximation,
though we present a more detailed analysis later.

The small value of $\zeta_N /\zeta$, as well as $\zeta$'s for other
particles with weaker interactions, might help to understand why the
exponent in $a_*(t_*)\sim t_*^{\alpha_*}$ for the dust-dominated
universe as given by [\cite{yf4}]
\beq
\alpha_* =\frac{1}{6}\left( 4-\frac{\bar{\zeta}}{\zeta} \right)
\label{qtst-21}
\eeq
is close to 2/3, where $\bar{\zeta}$ means an average of $\zeta$'s for particles comprising the dust matter.

The vacuum energy of matter fields is estimated roughly of the order
of $E_{ve}\sim M_{*sb}^4$  According to \reflef{qtst-19}) we may expect $E_{ve}\sim 10^{-24}$ at the earliest epoch.  Multiplying this by $e^{-4\zeta_q\sigma}$ will give another potential $V_1(\sigma)$.  Due to $\zeta_q < \zeta$, however, the new potential will soon overwhelm  $V(\sigma)$, eventually reproducing the excess by 60 orders of magnitude at the present epoch.  The ``vacuum energy problem" seems to call for a novel mechanism which is yet to be discovered.

We also add that the calculation leading to \reflef{qtst-18}) and
\reflef{qtst-19}) is closely related to the trace anomaly
[\cite{tranom}].  In fact the J frame Lagrangian with the matter part
\reflef{qtst-17}) but ignoring $\Lambda$, for the moment, is invariant
under global scale transformation (dilatation).  We find that $\sigma$
in E frame plays the role of the associated massless Nambu-Goldstone boson (dilaton).  Due to the quantum correction, the dilatation symmetry is ultimately broken explicitly, making $\sigma$ the pseudo Nambu-Goldstone boson which is massive.

%%%%%%%%%%%%%% sect 4 %%%%%%%%%%%%%%%%%%%%%%%%%%%%%%%%%%%%%%%%%%%%%%%%%%%%%
\section{Non-Newtonian gravity}

In E frame as an approximately physical CF,  we expect that the scalar
field $\sigma$ is decomposed into the sum of the cosmological background part $\sigma_b(t_*)$ and the locally fluctuating component $\sigma_f(x)$;
\beq
\sigma(x) = \sigma_b(t_*) + \sigma_f(x).
\label{qtst-22}
\eeq
Substituting this into \reflef{qtst-18}) and using \reflef{qtst-15})
with $\sigma$ replaced by $\sigma_b(t_{*{\rm p}})$ to give the  quark mass $m_{*q}$, we obtain
\beq
{\cal L}_{mqf} = \sqrt{-g_*}\zeta_q m_{*q}\bar{q}_*q_*\sigma_f,
\label{qtst-22a}
\eeq
which implies that $\sigma_f$  mediates a force among matter 
objects.

We may also apply a familiar field theoretic calculation 
obtaining the self-mass $\mu_f$ arising from a quark loop, for
example, as estimated to be
\beq
\mu_f^2 \sim m_{*q}^2 M_{*sb}^2.
\label{qtst-23}
\eeq
By using today's values for the masses, we obtain $\mu_f\sim 0.84\times
10^{-36}\sim 2.1 \times 10^{-9}{\rm eV}$, and the corresponding force
range $\lambda = \mu_f^{-1}\sim 1.2\times 10^{36}= 9.6 \times 10^3{\rm
cm}\approx 100{\rm m}$.  We should allow, however, a latitude of
several orders of magnitude due to ambiguities in evaluating the
self-energy.  It seems nevertheless unlikely that the force-range is
as larger as the size of the solar-system to allow the solar-system
experiments to constrain the coupling strength $\zeta$.

It has been argued, on the other hand, that the force mediated by $\sigma$ is
long-range, because the second derivative of $V(\sigma)$ is extremely
small [\cite{qt2}].  This can be derived first by noting that the left-hand side of \reflef{qtst-7}) is $\sim t_{*\rm p}^{-2}$ as long as the universe at
the present time $t_{*{\rm p}}$ expands according to a power law,
placing an upper bound on $V$ on the right-hand side.  If the potential is
sufficiently flat, like the exponential potential, for example, the
mass squared defined by the second derivative should be of the same
order of magnitude as $V$ itself.  The corresponding force-range is
$\sim t_{*{\rm p}}$, which is the size of the whole visible universe.

The squared mass given by \reflef{qtst-23}) is overwhelmingly larger than $\sim
t_{*\rm p}^{-2}$.  In this connection we  point out, however,  that the above argument for a long-range force
applies to $\sigma_b$, a classical background field supposed to obey a 
nonlinear equation, whereas \reflef{qtst-23}) is related to  the
solution of a linear  harmonic oscillator, which is the basis of
the concept of a quantum.  The two kinds of mass can be entirely different from
each other.   A well-known example is provided
by the sine-Gordon equation in 2-dimensions [\cite{sol}].  The
quantized field has ``mesonic excitations" at each of the sinusoidal
potential minima, with the mass squared given by the second derivative
of the potential.  Quite apart from them, there are classical soliton
solutions, which connect the potential minima.  The mass of each of such
solutions is in fact different from the mass of the
meson.

The analogy is far from complete in our model.  The squared mass $\sim t_*^{-2}$ has nothing to do with the soliton mass.  We even do
not know if there is a soliton-type solution of our cosmological
equation, although we might expect a mechanism of nonlinear dynamics
particularly for the effect of the vacuum energy that includes the
effect of the fully nonlinear extension of \reflef{qtst-18}).   What still
interests us is that the soliton solution shows the behavior entirely
different from the propagation of the mesonic excitation.  In the same
way we may anticipate  the slow evolution of $\sigma_b(t_*)$ instead
of the oscillatory behavior.    Without entering any further details at
this moment, we simply propose a conjecture that the fluctuating
component acquires a nonzero mass given by \reflef{qtst-23}) in a way compatible with the slow rolling of the background field.

Given the urgency of accommodating WEP violation in the prototype BD 
model  and the inevitability of solving the vacuum-energy problem, it
might be a unique consequence of a viable model of scalar-tensor
theories with a cosmological constant to have $\sigma_f$ showing
itself quintessentially as non-Newtonian gravity, or the fifth force
in its scalar version [\cite{fsb},\cite{yfnat}].

The phenomenological analyses can be made most conveniently in terms of the static potential between two nuclei $a$ and $b$;
\beq
V_{5ab}(r) =-\frac{G m_am_b}{r }\left(1 + \alpha_{5ab} e^{-r/\lambda}\right),
\label{qtst-24}
\eeq
where the coefficient $\alpha_{5ab}$ is given basically by the one between two nucleons;
\beq
\alpha_{5N} = 2\zeta_N^2.
\label{qtst-25}
\eeq
According to \reflef{qtst-20}) we have $\alpha_{5N} \sim 10^{-3}$ for $\zeta =1$, which seems already on the verge of an immediate exclusion [\cite{fsb}], though the conclusion might be premature in view of uncertainties in the force-range as well as estimating composition-dependence coming from the nuclear binding energies.  It seems nevertheless interesting to suggest a possible link between the cosmological constant problem and non-Newtonian gravity.

%%%%%%%%%%%%%%%%%%%%%%%%% sect 5 %%%%%%%%%%%%%%%%%%%%%%%%%%%%%%%%%%%%

\section{Discussions}

The $\Lambda$ term in \reflef{qtst-1}) may depend on $\phi$, as suggested by some examples of higher-dimensional theories.  In other words, we may allow self-interaction of the scalar field.   Let us consider a monomial $\phi^{\ell}\Lambda$, for simplicity.  After the conformal transformation \reflef{qtst-2}) we obtain
\beq
V(\sigma)=\Lambda\xi^{-\ell /2}e^{-4\zeta'\sigma},
\label{qtst-26}
\eeq
where
\beq
\zeta' = \left( 1-\frac{\ell}{4}\right)\zeta.
\label{qtst-27}
\eeq
Except for the choice $\ell = 4$, we have the same exponential
potential only with a different coefficient $\zeta'$.  The relation
\reflef{qtst-4}) remains the same as before.  The solution
\reflef{qtst-10})-\reflef{qtst-14}) are still correct if we replace
$\zeta$ everywhere by $\zeta'$ for the radiation-dominated universe,
while we encounter some complications for the dust-dominated universe,
because $\zeta$ on the right-hand sides of \reflef{qtst-8}) and
\reflef{qtst-9}) remain unchanged.  Likewise, the constant $\zeta$ that determines the matter coupling in \reflef{qtst-15}) and \reflef{qtst-19}) is still $\zeta$.  As a consequence, we now have $m_* \sim t_*^{-(\zeta/\zeta')/2}$.

In the previous discussions, the strength of the matter coupling is
constrained from below because $|\zeta| > 1/2$ from the physical
condition in \reflef{qtst-12}).  We can relax this constraint by
noting that the condition is now for $\zeta'$ and  \reflef{qtst-27}) allows $|\zeta| < |\zeta'|$ for $\ell <0$ or $\ell >8$.  This might make it easier for $\alpha_{5N}$ as given by \reflef{qtst-25}) to be consistent with experiments.  We may have even more flexibility if we allow a more general function of $\phi$ multiplied by $\Lambda$ in \reflef{qtst-1}).

For the prototype model, we showed that the universe is static in J frame.  This conclusion is subject to a change for a nonzero $\ell$.  We find $a(t)=t^{\alpha}$ with $\alpha =((1/2), \mbox{or} \hspace{.5em}(2/3))\times$\\$(\ell/(\ell -2))$ if $\ell \neq 2$, for the radiation- or dust-dominated universe, respectively, if $\ell \ne 2$, while $a(t)$ shows an exponential behavior if $\ell =2$ [\cite{yf4}].  In this way departing from WEP may not appear so much urgent.  However, obtaining $\alpha$ which agrees with  the standard value within the difference of $\pm$10\%, for example, requires either $\ell < -18$ or $\ell > 22$, somewhat extreme choices.  Also $\alpha$ is negative for $0<\ell <2$.

In Section 3, we were content with having E frame as an {\em approximately} physical CF.  Strictly speaking, however, we should move to another CF in which particle masses stay constant.  Let us do this by assuming again the exponential dependence for $\ell =0$;
\beq
m_{*N}(t_*)\sim t_*^{-\delta /2},
\label{qtst-28}
\eeq
where $\delta = \zeta_N /\zeta$.  Remember that without WEP masses of
different species of particles may behave differently.  The time
variable $d\tilde{t}$ measured in units of $m_{*N}^{-1}$ is defined,
analogously to \reflef{qtst-15a}), by  $d\tilde{t} =
dt_*/m_{*N}^{-1}$, yielding    
\beq
\tilde{t}\sim t_*^{1+\delta /2}.
\label{qtst-29}
\eeq
By using also the second relation in \reflef{qtst-15a})  we obtain
\beq
\tilde{a} \sim m_{*N} a_* \sim \tilde{t}^{\tilde{\alpha}},
\label{qtst-30}
\eeq
where the exponent
\beq
\tilde{\alpha}\approx\frac{2}{3}\left( 1-\frac{3}{2}\delta \right),
\label{qtst-31}
\eeq
has been derived from \reflef{qtst-21}) for the dust-dominated universe in E frame, keeping only terms proportional to $\delta \approx 0.02$ shown in \reflef{qtst-20}).

A small deviation from 2/3 as in \reflef{qtst-31}) may not be
seriously important.  We then try to show that the gravitational
constant is now time-dependent to the extent that it is close to the
observational upper bounds available so far.  For this purpose we
consider \reflef{qtst-18}) assumed to be exponentiated with $q$
replaced by $N$.  Under the  conformal transformation
$g_{*\mu\nu}\rightarrow \tilde{\Omega}^2 \tilde{g}_{\mu\nu}$, the mass
is transformed as $m_* \rightarrow \tilde{m} =m_*
\tilde{\Omega}^{-1}$.  We therefore choose $\tilde{\Omega}\sim
m_{*N}$.  Ignoring  terms of higher order in $\delta$, we obtain
\beq
\half \sqrt{-g_*}R_* \approx \half \sqrt{-\tilde{g}}\tilde{\Omega}^{-2}\tilde{R},
\label{qtst-32}
\eeq  
in which we identify the gravitational constant in the tildered CF as
\beq
8\pi \tilde{G}=\tilde{\Omega}^2 \sim t_*^{-\delta}.
\label{qtst-33}
\eeq
Further using $\tilde{t}\sim t_*$ obtained from \reflef{qtst-29}) by omitting the $\delta$-dependence, we finally find
\beq
\frac{\dot{\tilde{G}}}{\tilde{G}} \sim -\delta \tilde{t}^{-1},
\label{qtst-34}
\eeq
which can be compared with the observed upper bound $(0.2\pm
0.4)\times 10^{-11}{\rm y}^{-1}$ [\cite{vik}].  Further improving the
accuracy will test the proposed theoretical model.

Note that we have applied the conformal transformation only in the
context of the classical background field, leaving \reflef{qtst-22a})
still accepted as the coupling of the quantized field almost unaffected.

We add, however, that there is a theoretical model [\cite{yf3}] featuring the
scalar field that stays nearly constant for some time duration,
allowing us to understand an extra acceleration of the universe, as
indicated by recent observations.  During this period supposed to
cover the present epoch, we can avoid the issue of choosing a CF, also
predicting the time variation of the gravitational constant at the
level much lower than $10^{-10}{\rm y}^{-1}$.  We reasonably expect
 again that the matter coupling of $\sigma_f$ remaining nearly the same even
for the modified version of the model.

We point out, on the other hand, that it might take some time for the observational studies before the presence of the cosmological constant will be finally established.  We must still answer the question how the discrepancy of 120 orders of magnitude is avoided.  The issue of $\dot{G}/G$ is related to this part of the question.
\bigskip
\begin{center}
{\large\bf Acknowledgments}
\end{center}

I thank Naoshi Sugiyama, Akira Tomimatsu and Kei-ichi Maeda for many
valuable discussions.  Conversations with Yoshio Ohnuki and Shinsaku
Kitakado are also appreciated.
\bigskip

\begin{center}
{\large\bf References}
\end{center}
\begin{enumerate}

%%%%%%%%%%%%%%%%%%%%%%%%
\item\label{qt1}T. Chiba, N. Sugiyama and T. Nakamura, Mon. Not. R. Astron. Soc., {\bf 289}, L5(1997); R.R. Caldwell, R. Dave and P.J. Steinhardt, Phys. Rev. Lett. {\bf 80}, 1582(1998); W. Hu, D.J. Eisenstein and M. Tegmark, Phys. Rev. {\bf D 59}, 023512(1998); G. Huey, L. Wang, R.R. Caldwell and P.J. Steinhardt, Phys.Rev. {\bf D59}, 063005(1999); P.G. Ferreria and M. Joyce, Phys. Rev. Lett. {\bf 79}, 4740(1997); Phys. Rev. {\bf D58}, 023503(1998); F. Rosati, hep-ph/9906427; J. Frieman and I. Waga, Phys. Rev. {\bf D57}, 4642(1998); P.J. Steinhardt, L. Wang and I. Zlatev, Phys. Rev. {\bf 59}, 123504(1999); M.C.Bento and O. Bertolami, gr-qc/9905075; F. Perrotta, C. Baccigalup and S. Matarrese, astro-ph/9906066.
%%%%%%%%%%%%%%%%%%%%%%%%
\item\label{qt2}S.M. Carroll, astro-ph/9806099; T. Chiba,
gr-qc/9903094; N. Bartolo and M. Pietroni, hep-ph/9908521.

%%%%%%%%%%%%%%%%%%%%%%
\item\label{sn}P. Garnavich {\it et al.}, Astrophys J., {\bf 493}, L3(1998); S. Perlmutter {\it et al.}, Nature {\bf 391}, 51(1998); Astrophys J., astro-ph/9812133.
\item\label{ccp}A.D. Dolgov, {\sl The very early universe}, Proceedings of Nuffield Workshop, 1982, eds. B.W. Gibbons and S.T. Siklos, Cambridge University Press, 1982; Y. Fujii and T. Nishioka, Phys. Rev. {\bf D42}, 361(1990).

%%%%%%%%%%%%%%%%%%%%%%
\item\label{yf3}Y. Fujii and T. Nishioka, Phys. Lett. {\bf B254}, 347(1991); Y. Fujii, Astropart Phys. {\bf 5}, 133(1996); Y. Fujii, gr-qc/9908021.
%%%%%%%%%%%%%%%%%%%%%%%
\item\label{yf4}Y. Fujii, Prog. Theor. Phys. {\bf 99}, 599(1998).
%%%%%%%%%%%%%%%%%%%%%%%
\item\label{bd}C. Brans and R.H. Dicke, Phys. Rev. {\bf 124}, 925(1961).
%%%%%%%%%%%%%%%%%%%%%%%
\item\label{fsb}See, for example, E. Fischbach and C.L. Talmadge, {\sl The search for non-Newtonian gravity}, AIP Press-Springer, N.Y., 1998.
%%%%%%%%%%%%%%%%%%%%%%%
\item\label{tranom}M.S. Chanowitz and J. Ellis, Phys. Lett. {\bf 40B},
397(1972); R.D. Peccei, J. Sola and C. Wetterich, Phys. Lett. {\bf B195}, 183(1987).
%%%%%%%%%%%%%%%%%%%%%%%
\item\label{sol}See, for example, S.V. Ketov, Fortsch. Phys. {\bf 45},
237(1997), hep-th/9611209.
%%%%%%%%%%%%%%%%%%%%%%%
\item\label{yfnat}Y. Fujii, Nature Phys. Sci. {\bf 234}, 5(1971); Int. J. Mod. Phys. {\bf A6}, 3505(1991).
%%%%%%%%%%%%%%%%%%%%%%%
\item\label{vik}R.W. Hellings, {\it et al.}, Phys. Rev. Lett. {\bf 51}, 1609(1983).
\end{enumerate}
%%%%%%%%%%%%%%%%%%%%%%%%%%%%%%%%%%%%%%%%%%%%%%%%%%%%%%%%%%%%%%%%%%%%%%%%%%
%%%%%%%%%%%%%%%%%%%%%%%%%%%%%%%%%%%%%%%%%%%%%%%%%%%%%%%%%%%%%%%%%%%%%%%%%%
\begin{figure}[h]
\hspace*{1.cm}
\epsfxsize=14.5cm
\epsffile{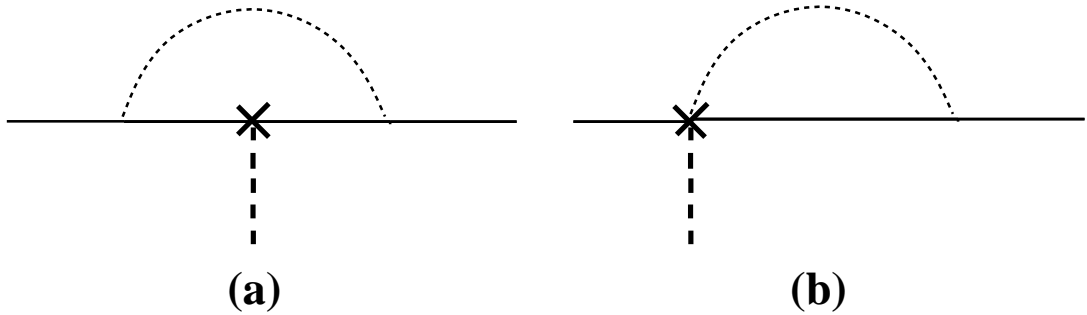}
\caption{One-loop diagrams contributing to the anomaly, in the
framework of $N$-dimensional regularization.  We started with
spacetime dimensionality $N\neq 4$, which renders loop integrals 
expressible in temrs of Gamma functions $\Gamma (2-N/2)$.  On the
other hand, vertices denoted by crosses, the mass term in (a) while
the interaction term in (b), are proportional to $N-4$,
which multiplies with the pole in the gamma function to produce a
finite result [\cite{yf4}].  The solid and dotted lines represent the
quark and the gluon field, respectively, while the dashed line is for $\sigma$. }
\label{fq1} 
\end{figure}

\end{document}